\documentclass[doublecol]{epl2} 
\usepackage{bm}
\usepackage{amsmath}
\newcommand{\ox}{\overline{x}}
\newcommand{\oy}{\overline{y}}

\newcommand{\pdiff}[2]{\frac{\partial #1}{\partial #2}}

\newcommand{\be}{\begin{equation}}
\newcommand{\ee}{\end{equation}}
\newcommand{\ba}{\begin{eqnarray*}}
\newcommand{\ea}{\end{eqnarray*}}
\newcommand{\tr}{{\rm Tr}}
\newcommand{\new}{\nonumber\\}
\newcommand{\abs}[1]{\left|#1\right|}

\newcommand{\erf}{{\rm erf}}

\newcommand{\hA}{\hat{A}}
\newcommand{\hvar}{\hat{\varphi}}

\title{One-dimensional Kac model  of dense amorphous hard spheres}
\author{H. Ikeda\inst{1} \and A. Ikeda\inst{2}}
\shortauthor{H. Ikeda \etal}

\institute{ 
\inst{1} Department of Physics, Nagoya University - Nagoya, 464-8602, Japan\\ 
\inst{2} Fukui Institute for Fundamental Chemistry, Kyoto University - 
Takano-Nishihiraki-cho 34-4, Sakyo-ku, Kyoto, 606-8103, Japan 
}

\abstract{ We introduce a new model of hard spheres under confinement
for the study of the glass and jamming transitions.  The model is an
one-dimensional chain of the $d$-dimensional boxes each of which
contains the same number of hard spheres, and the particles in the boxes
of the ends of the chain are quenched at their equilibrium positions.
We focus on the infinite dimensional limit ($d \to \infty$) of the model
and analytically compute the glass transition densities using the
replica liquid theory.  From the chain length dependence of the
transition densities, we extract the characteristic length scales at the
glass transition. The divergence of the lengths are characterized by the
two exponents, $-1/4$ for the dynamical transition and $-1$ for the
ideal glass transition, which are consistent with those of the $p$-spin
mean-field spin glass model.  We also show that the model is useful for
the study of the growing length scale at the jamming transition.  }

\begin{document}

\maketitle

\section{Introduction}

When liquids are rapidly cooled well below the melting temperature, the
relaxation time and viscosity drastically increase and eventually they
freeze into the disordered state.  This phenomenon is known as the glass
transition~\cite{ Debenedetti2001}.  The mechanism of this slowing down
and the presence/absence of an underlying genuine phase transition are
under active debate~\cite{
gotze2008complex,Berthier2011RMP,tanaka2010critical}.

A good way to understand theoretically the glass transition would be to
first construct a mean field theory and then consider finite dimensional
effects. Recently the glass transition of the infinite dimensional hard
spheres is studied as a mean-field model, and the exact thermodynamic
theory of the model is developing~\cite{
parisi2010mean,kurchan2012exact,kurchan2013exact,
charbonneau2014exact}. The glass transition of the model can be
characterized by the two relevant densities~\cite{parisi2010mean}.  The
first is the dynamical transition density, $\varphi_d$, below which the
phase space is divided into an exponentially large number of metastable
states.  In the mean-field limit, the dynamics of the system gets frozen
at this density as the system is trapped into one of the metastable
states.  With increasing the density, the number of the metastable
states decreases and eventually becomes sub-exponential, at which the
system undergoes the thermodynamic transition into the ideal glass
state.  This transition density $\varphi_K$ is called the Kauzman
density.

What types of finite dimensional effects play a dominant role in the
glass transition is not clear yet~\cite{Berthier2011RMP,biroli2013jcp}. One promising
theoretical scenario which incorporates an effect of fluctuations
of finite dimensional systems is proposed as the Random First Order
Transition (RFOT) theory~\cite{kirkpatrick1989scaling}.  This theory
assumes that the divergence of the relaxation time at $\varphi_d$ is
avoided, and at $\varphi$ between $\varphi_d$ and $\varphi_K$, 
the liquid state can be seen as a patchwork of the local metastable
configurations~\cite{bouchaud2004adam}.  The theory predicts that the
characteristic size of these domains increases with increasing the
density and eventually diverges at $\varphi_K$, where the
thermodynamic glass transition occurs.

In practice, this length can be measured through the thermodynamic behavior of 
liquids under confinement~\cite{bouchaud2004adam}.
We prepare an equilibrium configuration of particles and pin only the particles outside
of a cavity of the size $L$.
One can define and compute the critical cavity size at
which the boundary affects the thermodynamics of the system in the
cavity, which is referred to as the point-to-set (PS) correlation
length.  Numerical simulations of realistic liquid models under 
this confinement have been carried out~\cite{
biroli2008thermodynamic,kob2012non,glen2012}, which confirmed that the
PS length indeed grows as the systems slow down.  However, computing
theoretically the PS length of realistic liquids is still very
challenging~\cite{franz2013static,cammarota2014first}.  One of the
difficulties is due to the inhomogeneous nature of the confined liquids,
which requires the inhomogeneous version of the liquid state
theory~\cite{biroli2006inhomogeneous}.  The $p$-spin mean-field
spin glass model under confinement has been extensively
studied~\cite{franz2007analytic,cammarota2013confinement}.  The studies
showed that the PS length of the model diverges at the Kauzuman
transition point with the power-law behavior with the exponent $-1$ and
another length characterizing the local stability of the metastable
states diverges at the dynamical transition point with the exponent
$-1/4$~\cite{franz2007analytic}.

The main purpose of the present letter is to propose a new model of hard
spheres under confinement, which can be analyzed using the {\em
homogeneous} liquid state theory~\cite{hansen1990theory}.  Focusing on
the large dimensional limit of the model, we compute analytically the PS
length using the replica liquid theory.

Since the model proposed is not a spin model but hard spheres, the model
undergoes another type of the phase transition, called the jamming
transition, at the higher
density~\cite{parisi2010mean,torquato2000random}. This is the transition
of the liquid state into the state where hard spheres are packed so
closely that the pressure diverges.  We additionally show that the model
can be used to study the growing lengths at the jamming transition.

\section{Model}

\begin{figure}
\begin{center}
\onefigure[width=8cm]{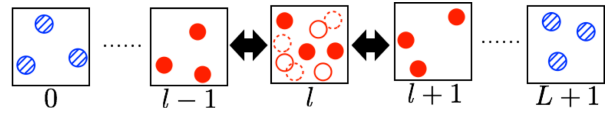} \caption{ A typical configuration of
the model: The position of the particles in each boxes labeled $0$-th to $L+1$-th
are schematically shown as cycles.  The shaded cycles represent the
frozen particles in $0$-th and $(L+1)$-th boxes, and the filled cycles
represent the mobile particles in each box.  In the $l$-th box, the
particle positions in the $(l-1)$-th and $(l+1)$-th box are also shown as
the solid and dashed cycles respectively to indicate the particles in
$l$-th box can not overlap with the particles in $(l-1)$-th and $(l+1)$-th
boxes.  } \label{fig.1}
\end{center}
\end{figure}

Consider an one-dimensional chain of $L+2$ boxes.  Each box is a
$d$-dimensional cube of the volume $V$ and contains $N$ hard spheres.
We assume that a particle interacts with the other particles in the same
box and those in the nearest two boxes.  Therefore, the Hamiltonian of
the model is
\begin{equation}
H = \sum_{l=0}^{L+1} \sum_{i<j}^{N}v(\abs{x_i^{l}-x_j^{l}}) +
\sum_{l=0}^{L}\sum_{ij}^Nv(\abs{x_i^{l}-x_j^{l+1}})\label{163322_2Dec14},
\end{equation}
where the $d$-dimensional vector $x_i^l$ denotes the position of the
$i$-th particle in the $l$-th box and $v(r)$ is the interaction potential 
between hard spheres of the diameter $D$, hence $v(r)=\infty \ (r\leq D),\ 0 \ (r>D)$.  
We study the model under the condition that the particles in the
$0$-th and $(L+1)$-th boxes are quenched at their equilibrium positions.
Thus the model is interpreted as the $d$-dimensional hard spheres
confined by the amorphous walls. We sketch a typical configuration of
our model schematically in Fig.~\ref{fig.1}.

As it is clear from the Hamiltonian, eq.~(\ref{163322_2Dec14}), we can
treat the model as a $L+2$ component {\em homogeneous} liquid,
which enables us to analyze the model using the {\em homogeneous} liquid
state theory.  This is one of the advantages of this model. In
particular, the free energy of this model can be evaluated analytically
at the large dimensional limit using the replica liquid theory. 
In the following sections, we calculate the two relevant densities 
$\varphi_{d}$ and $\varphi_{K}$ for the
glass transition as a function of the number of the boxes $L$.  By
interpreting $L$ as the correlation length, we extract the lengths $\xi_{d}$
and $\xi_{K}$ which diverge at $\varphi_d$ and $\varphi_K$,
respectively.  Additionally, we also apply this method to the jamming
transition to discuss the possibility to compute its growing length
scales.

\section{Free energy at large dimensional limit}

To compute the free energy of the model, we employ the replica liquid
theory assuming the one-step replica symmetry breaking (1RSB) ansatz~\cite{
mezard.parisi,monasson1995structural,parisi2010mean}.  The
main idea of the theory is to consider $m$ copies (replicas) of the
original system. There is no interaction between the replicas, and
therefore the partition function of the replicated system is written as
\begin{align}
Z_m &\equiv\prod_{a=1}^{m}\tr_{x_{a}} e^{-
\sum_{a=1}^mH[\left\{ x_a \right\}]/T},\label{023601_31Dec14}
\end{align}
where $H$ is the Hamiltonian of the original system, and $T$ is the
temperature. From now on, we set $T=1$ because the temperature is an
irrelevant variable for hard spheres. Within the replica theory, the
logarithm of the number of the minima, which is called the configurational entropy, $\Sigma$, can
be calculated by the formula~\cite{monasson1995structural}:
\begin{align} 
 \Sigma=-\frac{m^2}{N}\pdiff{}{m}\left(\frac{\log Z_m}{m}\right).
\label{154904_14Mar15}
\end{align}

If one takes the large dimensional limit ($d\to\infty$) and the thermodynamic limit
($N\to\infty$, $V\to\infty$, $N/V$=const), the calculation
of the partition function of the replicated system can be greatly
simplified because the higher order terms of the Mayer cluster expansion
become negligible, and only the first term needs to be
considered~\cite{hansen1990theory,parisi2010mean}.  The replicated
free-energy of our model is
\begin{align}
\log Z_{m} &= \sum_{l=1}^L\int d\ox
\rho_{l}(\ox)\left(1-\log\rho_{l}(\ox)\right)\new &+
\frac{1}{2}\sum_{l=1}^{L}\int d\ox
d\oy\rho_l(\ox)\rho_l(\oy)f(\ox-\oy)\new &+ \sum_{l=1}^{L-1}\int d\ox
d\oy \rho_l(\ox)\rho_{l+1}(\oy)f(\ox-\oy)\new & + \int d\ox
dy\rho_{1}(\ox)\hat{\rho}_{0}(y)f(\ox-y)\new &+ \int d\ox
dy\rho_{L}(\ox)\hat{\rho}_{L+1}(y)f(\ox-y)\label{032629_13Oct14},
\end{align}
where $\ox=\left\{ x^1,\cdots x^m \right\}$ denotes the positions of
hard spheres in the replicated space, and $\rho_l(\ox)= \left\langle
\sum_i\prod_{a=1}^{m}\delta(x^{a}-x_i^{l,a}) \right\rangle$ is the
density distribution of the $l$-th box.  The boundary conditions are
included in the last two lines of eq.~(\ref{032629_13Oct14}), where
$\hat{\rho}_0(x)=\sum_{i}\delta(x-x_{i}^{0})$ and
$\hat{\rho}_{L+1}(x)=\sum_{i}\delta(x-x_{i}^{L+1})$ are the microscopic
density distributions of the frozen particles belonging in the $0$-th and
$(L+1)$-th boxes.  $x_i^0$ and $x_i^{L+1}$ are taken from their
equilibrium positions. 
If the system is sufficiently large, the
self-averaging properties for the free-energy holds:
\begin{align}
\log Z_m\approx \overline{ \log Z_m},
\end{align}
where the overline denotes the average of 
$\hat{\rho}_0(x)$ and
$\hat{\rho}_{L+1}(x)$. Since the free-energy eq.~(\ref{032629_13Oct14})
depends linearly on those variables, we have only to replace as
$\hat{\rho}_0(x)\rightarrow \rho$ and $\hat{\rho}_{L+1}(x)\rightarrow
\rho$, where $\rho=N/V$ is the number density of each boxes.

In the next step, we introduce the 1RSB Gaussian
ansatz~\cite{parisi2010mean}:
\begin{align}
\rho_l(\ox) &= \rho\int
dX\prod_{a=1}^{m}\gamma_{A_l}(x_a-X), \label{143017_21Jan15}
\end{align}
where $\gamma_A(x) = \exp\left(-x^2/2A\right)/(2\pi A)^{d/2}$.  This
ansatz claims that the distribution of the replicated particles is 
the Gaussian with the variance $A_{l}=\left\langle
\sum_{a<b}(x^{l,a}-x^{l,b} \right\rangle/m(m-1)$. Substituting
eq.~(\ref{143017_21Jan15}) into eq.~(\ref{032629_13Oct14}) and taking
the large dimensional limit~\cite{parisi2010mean}, one obtains the analytical
expression of the free energy:
\begin{align}
\frac{\log Z_m}{N} &= S_{id}+S_{int},\label{023057_31Dec14}\new
S_{id} &= \sum_{l=1}^L
\frac{d}{2} \left[ (m-1)\log \frac{\hA_{l}}{d^2}+ \log \frac{m}{d}+m
\right],\new S_{int} &= -\frac{3L\hvar}{2} \left( 1+ \frac{2}{3L} \right)
 + \frac{\hvar}{2}\sum_{l=1}^L G_m ( \hA_l )\new &+ \hvar\sum_{l=0}^{L}G_m
\left( \frac{\hA_{l}+\hA_{l+1}}{2} \right),
\end{align}
where we introduced the normalized volume fraction $\hvar=2^d\varphi/d$ and the
normalized cage size $\hA_l = A_l d^2/D^2$. $G_m(\hA)$ is the auxiliary function given by
\begin{align}
G_{m}(\hA) &= \int_{-\infty}^{\infty}dy e^y \left[ \Theta \left(
\frac{y+\hA}{\sqrt{4\hA}} \right)^m-\theta(y) \right],\label{182338_2May15}
\end{align}
where $\Theta(x) = \left[ \erf(x)+1 \right]/2$.  In
eq.~(\ref{023057_31Dec14}), the cage size of
the $0$-th and $(L+1)$-th boxes are zero, $\hA_0=\hA_{L+1}=0$, because of
the boundary condition. $\hA_l$ for other $l$ is calculated from the saddle point
conditions for $\hA_l$:
\begin{align}
\frac{1}{\hA_l}
&= \frac{2}{(1-m)d}\pdiff{S_{int}}{\hA_l}, &(l=1, \ldots, L).
\label{220113_1Jan15}
\end{align}
Substituting the solution of this set of equations into
eq.~(\ref{023057_31Dec14}), the free energy of the $m$ replicated system
is obtained.

\section{Glass transition}
Here, we evaluate the correlation length of the glass transition.
As mentioned before, there are the two relevant
densities for the glass transition, $\hvar_d$ and $\hvar_K$. Their values
can be computed by analyzing the free energy at $m=1$.

\begin{figure}
\begin{center}
\onefigure[width=8cm]{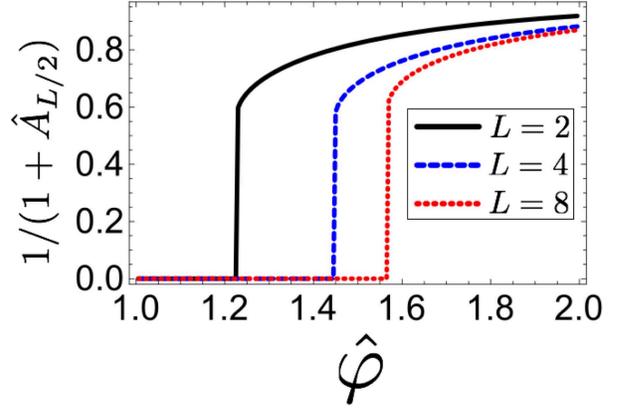} \caption{The density dependence of the cage
size for several $L$'s.  The solid line is for $L=2$, the dashed line is
for $L=4$ and the dotted line is for $L=8$.}  \label{192840_20Jan15}
\end{center}
\end{figure}

First, we examine the behavior of our model near the dynamical
transition point, $\hvar_d$, where the exponentially many metastable
states emerge on the free energy. In the framework of the replica liquid
theory, the order parameter to characterize the metastable states is the
cage size, which is given by the solutions of
eq.~(\ref{220113_1Jan15})~\cite{parisi2010mean}.  Their numerical
solution for $\hA_l$ of the $L/2$-th box, which is the farthest from the
boundaries, are shown in Fig.~\ref{192840_20Jan15}. For the small densities,
the cage size $\hA_{L/2}$ is infinity ($1/(1+\hA_{L/2})=0$), meaning that
the system is ergodic and in the liquid phase. Increasing the
density, the cage size jumps discontinuously to a finite value
at the dynamic transition point, $\hvar_d$.

\begin{figure}
\begin{center}
\onefigure[width=8cm]{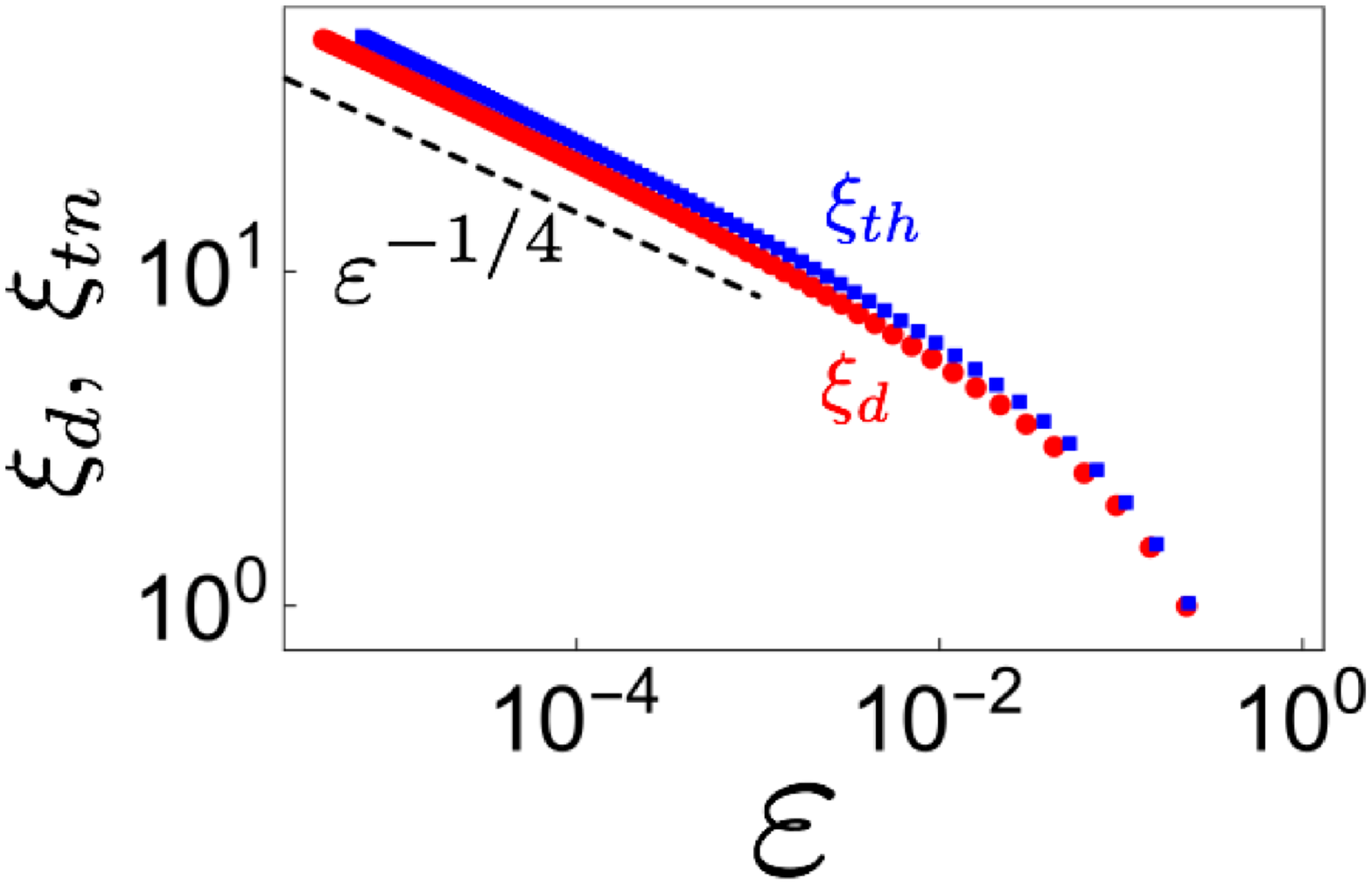} \caption{ The correlation legnths near
the $\hvar_{d}$ and $\hvar_{th}$. The filled cycles are for $\xi_d$ as a
function of $\varepsilon=(\hvar_{d}-\hvar)/\hvar_d$. The filled squares are for
$\xi_{th}$ as a function of $\varepsilon=(\hvar_{th}-\hvar)/\hvar_{th}$. }
\label{222953_1Jan15}
\end{center}
\end{figure}

From Fig.~\ref{192840_20Jan15}, it is clear that $\hvar_d$
increases as increasing $L$.
From this result, we can convert $\hvar_d(L)$ to $L(\hvar)$,
the characteristic number of the
boxes as a function of the transition density. We define
the correlation length by $\xi_d = L(\hvar)/2$ and plot it in
Fig.~\ref{222953_1Jan15}\cite{franz2007analytic}.  We find that $\xi_d$
diverges at $\hvar_d$ as $\xi_d \approx (\hvar_d - \hvar)^{-1/4}$.

Next, we evaluate the correlation length near the Kauzman density by
calculating the configurational entropy, $\Sigma$, which is obtained
by plugging the free energy into eq.~(\ref{154904_14Mar15}).
The final expression in the large dimensional limit becomes
\begin{equation}
\Sigma(m,\hvar,L) = \left[ \frac{d}{2}\log d - \frac{3\hvar}{2}\left( 1+
\frac{2}{3L} \right) \right] +O(d)\label{025551_15Jan15},
\end{equation}
up to the order of $O(d\log d)$.  From this
expression, it is clear that $\Sigma$ vanishes at
\begin{align}
\hvar_K(L) = \frac{d\log d}{3(1+2/3L)}. 
\end{align}
Following the same argument as for $\xi_d$, the growing length around
the Kauzman density is
\begin{equation}
\xi_{K}=\xi_{GCP} = \frac{L}{2} =
\frac{\hvar}{3(\hvar_{K}-\hvar)}\label{025557_15Jan15},
\end{equation}
where $\hvar_{K}=d\log d/3$ is the Kauzman density of the bulk system
($L\to\infty$).  

\section{Jammintg transition}

Hard spheres also undergo the jamming transition when the
system is compressed quickly or making the pressure
infinity~\cite{torquato2000random}.  The infinite dimensional hard
spheres serve as a mean-field model of the jamming transition as well and are
studied extensively using the replica liquid theory.  The theory showed
that the jamming transition is also characterized by the two relevant
densities~\cite{parisi2010mean}.  The jamming state obtained by
compressing the ideal glass state up to the infinite pressure is
referred to as the glass close packing, and its density is denoted as
$\hvar_{GCP}$.  If the jammed stated is prepared by a fast compression
of low density hard spheres, the pressure becomes infinity at a much lower
density than $\hvar_{GCP}$\cite{chaudhuri2010jamming}.  The lowest
density of the jammed states $\hvar_{th}$ is called the threshold
density.

In this section, we show that the growing length scales at the jamming transition 
can be computed from the analysis of the jamming transition of the present model. 
We essentially follow the strategy developed in the previous section. 
Using the replica liquid theory with the 1RSB ansatz,  
we compute $\varphi_{th}$ and
$\varphi_{GCP}$ of the model as a function of the number of the boxes,
$L$, which naturally gives rise to the two characteristic lengths $\xi_{th}(\hvar)$
and $\xi_{GCP}(\hvar)$. 
Note that recent studies showed that 
the full RSB ansatz is needed for the fully exact computation 
near the jamming transition~\cite{kurchan2013exact,charbonneau2014exact}. 
However such a computation for the present model seems quite involved. 
Here, we wish to stick to the 1RSB ansatz
and demonstrate that the replica theory analysis of the model provides 
the analytical expressions of the growing length scales at the jamming transition. 

In the replica liquid theory, the divergence of the pressure corresponds
to take the $m\to 0$ limit, because $m$ is inversely proportional to the
pressure~\cite{parisi2010mean}. Thus $\hvar_{th}$ can be calculated by
taking the $m\to 0$ limit in the self-consistent equation for the order
parameter eq.~(\ref{220113_1Jan15}).  Since the cage size, $\hA_l$,
vanishes at the $m\to 0$ limit, we set
$\hA_l=m\alpha_l$\cite{parisi2010mean}.  This is substituted to
eq.~(\ref{220113_1Jan15}) before the $m\to 0$ limit is taken.  We
numerically solve the equation and find that the behavior of $\alpha_{L/2}$
is qualitatively the same as that of $A_{L/2}$ for the glass
transition (Fig.~\ref{192840_20Jan15}).  At low densities,
$\alpha_{L/2}=\infty$ is the only solution of eq.~(\ref{220113_1Jan15}).
This means that there are no jammed states at those densities.  At higher
densities, the solution of eq.~(\ref{220113_1Jan15}) becomes
finite. This transition density is the lowest density, $\hvar_{th}$, of
the jammed states.  As in the case of the glass transition, the
characteristic length $\xi_{th}(\hvar)$ is computed from the $L$
dependence of $\hvar_{th}$.  This length $\xi_{th}$ is also plotted in
Fig.~\ref{222953_1Jan15}.  We find the power-law divergence $\xi_{th}
\approx (\hvar_{th} - \hvar)^{-1/4}$.  Likewise, we analyze the
correlation length near $\hvar_{GCP}$. The logarithm of the number of the jammed states is
calculated by the $m\to 0$ limit of the configurational entropy
eq.~(\ref{025551_15Jan15}) and $\hvar_{GCP}$ is defined as the density
where the configurational entropy becomes zero.  From
eq.~(\ref{025551_15Jan15}), it is clear that the behavior of
$\hvar_{GCP}$ should be the same as those of $\hvar_K$ because
eq.~(\ref{025551_15Jan15}) is independent from $m$. Thus we obtain the
critical behavior of the diverging length near $\hvar_{GCP}$ as
$\xi_{GCP} \approx (\hvar_{GCP} - \hvar)^{-1}$.

\section{Summary and discussion}

In this letter, we considered an one-dimensional chain of the
$d$-dimensional boxes each of which contains $N$ hard spheres as a model
system of liquids under confinement. By focusing on the large
dimensional limit, we analytically computed the phase
diagram of the model.  From the chain length dependence of the
transition densities, we derived the critical behavior of the two relevant
length scales of the glass transition.  We also showed that there are the two
relevant length scales of the jamming transition, whose critical behaviors are
similar to those of the glass transition as long as the 1RSB ansatz is
assumed.  One of the advantage of
the model is that, although the system is under confinement, it can
be regarded as a homogeneous liquid, and thus the model can be analyzed
using the usual replica liquid theory.

\begin{figure}
\begin{center}
\onefigure[width=8cm]{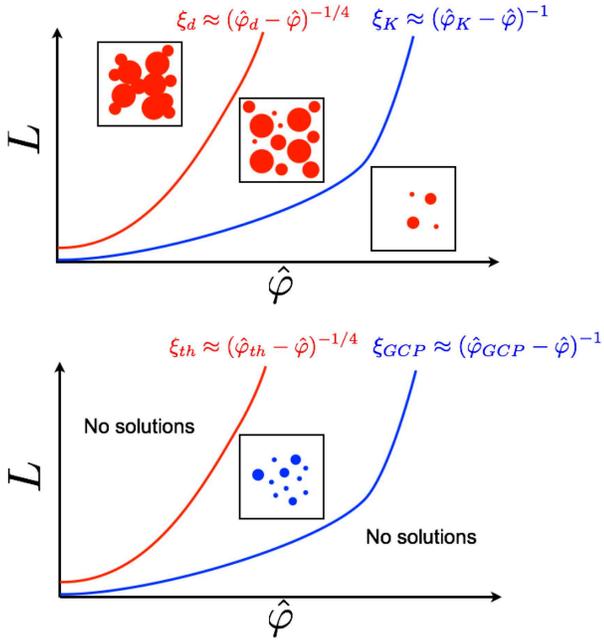} \caption{ The schematic phase
diagrams of our model: The upper and lower panel is the phase diagram for
the glass and jamming transition, respectively.  $\hvar =2^d\varphi/d$ 
is the normalized volume fraction and $L$ is the confinement
length. The insets are schematic pictures of the phase space in each
phase, where the cycles represent the accessible regions in the phase
space. } \label{071636_22Jan15}
\end{center}
\end{figure}

We summarize the results in Fig.~\ref{071636_22Jan15}.  The upper panel
is the phase diagram of the glass transition of the confined liquid,
where $\hvar$ is the normalized density $\hvar=2^d\varphi/d$, and $L$ is
the distance between the boundaries.  There are the two relevant length
scales $\xi_{d}$ and $\xi_K$, which diverge as $\xi_{d} \approx (\hvar_d
- \hvar)^{-1/4}$ and $\xi_{K} \approx (\hvar_K - \hvar)^{-1}$,
respectively. When $L>\xi_{d}$, the system holds the ergodicity and is
in the liquid phase. For $\xi_{K}<L<\xi_{d}$, the phase space splits into
the many sub-spaces and the ergodicity of the system is broken at least
in the large dimensional limit.  In $L<\xi_K$ region, the number of the
sub-spaces becomes sub-exponential and no longer contributes to the
entropy.  Thus the equilibrium phase transition from the liquid phase to the ideal
glass phase occurs.  These results are consistent with the results for the
$p$-spin mean-field spin glass
model~\cite{franz2007analytic,cammarota2013confinement}, where the
critical behaviors $\xi_{d} \approx (T - T_d)^{-1/4}$ and $\xi_{K}
\approx (T - T_K)^{-1}$ are observed.  This confirms that the proposed
model is considered to be a hard sphere extension of the $p$-spin
mean-field spin glass model under confinement.  The result for
$\xi_{d}$ is also consistent with the prediction of the inhomogeneous
mode-coupling theory for the dynamic correlation length diverging at the
mode-coupling transition point~\cite{biroli2006inhomogeneous}.

The lower panel of Fig.~\ref{071636_22Jan15} is the phase diagram of the
jamming transition.  There are the two relevant length scales $\xi_{th}$ and
$\xi_{GCP}$ which diverge as $\xi_{th} \approx (\hvar_{th} -
\hvar)^{-1/4}$ and $\xi_{GCP} \approx (\hvar_{GCP} - \hvar)^{-1}$,
respectively. When $L>\xi_{th}$, there are no jammed states. For
$\xi_{GCP}<L<\xi_{th}$, there exist exponentially many jammed
states. The number of the jammed states decreases with $L$, and becomes
sub-exponential when $L<\xi_{GCP}$.  

We should emphasize that the computation of $\xi_{th}$ and $\xi_{GCP}$ in this work 
is not exact, since the 1RSB ansatz is recently found to give unstable solution 
near the $m\to 0$ limit~\cite{kurchan2013exact}.  
If we use a better ansatz, the critical exponents may change~\cite{charbonneau2014exact}.  
Regarding this point, we wish to indicate the two points: 
(1) Even though the solution is unstable, the physics of
the 1RSB solution can still survive in certain situations~\cite{yoshino2014}
and thus the analysis given here is still useful~\footnote{
For examples, if the system is not quenched well into a single inherent
state during the compression and explores several different inherent
states, the physics of the 1RSB solution may appear~\cite{yoshino2014}.}.
(2) In case of $\xi_{GCP}$, the results would not be effected by this
instability at least in the large dimensional limit, because the expression
of the configurational entropy is $m$ independent. 

Finally we discuss the physical meaning of $\xi_{th}$ and $\xi_{GCP}$.
Several different types of lengths are known to diverge at the jamming
transition and the relations between the lengths are still not clear
enough~\cite{o2003jamming,silbert2005,wyart2005,
Ellenbroek2006,Wyart2010,hecke2010,teitel2011,
ikeda2013dynamic,Goodrich2013,Lerner2014,DeGiuli2014}.  Since $\xi_{th}$
and $\xi_{GCP}$ represent the distance between boundaries at which the
jammed configurations first appear/disappear, we expect that these
lengths correspond to the length detected in the finite size scaling
analysis $\xi_{FS} \sim |\varphi -
\varphi_J|^{-1.09}$~\cite{teitel2011}. However this is a tentative
expectation, because we applied a specific boundary condition (amorphous
boundary) which is suitable to compute the PS length at the glass
transition.  It is not clear which length should be detected by this
condition for the jamming transition.  To resolve this question, it
should be useful to study the present model under various different
types of confinements such as the flat walls~\cite{desmond2009random} or
random boundary~\cite{cammarota2013confinement}.

\acknowledgments We thank K. Miyazaki, H. Yoshino, G. Biroli and P. Urbani for helpful
discussions. HI acknowledge the JSPS Core-to-Core program and Program
for Leading Graduate Schools ``Integrative Graduate Education and
Research in Green Natural Sciences'', MEXT, Japan. AI acknowledges JSPS
KAKENHI No. 26887021.

\end{document}